# Pan-genome Analysis of Plastomes from Lamiales using PGR-TK

Aadhavan Veerendra[1] and Manoj Samanta[2,3,4]

1. Academy for Science & Design High School, Nashua, NH 03063.
2. Systemix Institute, Sammamish, WA 98074.
3. Coding for Medicine LLC, Sammamish, WA 98074.
4. samanta@homolog.us

# Abstract

Chloroplast sequences from the Lamiales order were analyzed using the Pangenome Research Toolkit (PGR-TK). Overall, most genera and families exhibited a high degree of sequence uniformity. However, at the genus level, *Utricularia*, *Incarvillea*, and *Orobanche* stood out as particularly divergent. At the family level, Orobanchaceae, Bignoniaceae and Lentibulariaceae displayed notably complex patterns in the generated plots. The PGR-TK algorithm successfully distinguished most genera within their respective families and often recognized misclassified plants.

# Introduction

The Lamiales is a large and diverse order of flowering plants within the asterid clade of eudicots, encompassing approximately 24,000 species [1]. Plants in this order vary widely, ranging from small herbs to shrubs and trees, and demonstrate a remarkable array of ecological adaptations and specialized traits. Notable examples include (i) the Lamiaceae family, known for its aromatic oils; (ii) the Oleaceae family, which thrives in Mediterranean climates; (iii) desiccation-tolerant "resurrection plants"; (iv) numerous carnivorous species, such as *Utricularia* and *Genlisea*; and (v) parasitic plants from the Orobanchaceae family. Several economically important plants, such as teak, olive, lavender, jasmine, and many culinary herbs (mint, basil, sage, and rosemary) are members of this order.

Molecular phylogenetic studies of the Lamiales order currently recognize 24-26 families [2-5]. Prominent families include Lamiaceae (mint family), Oleaceae (olive family), Verbenaceae (vervain family), Plantaginaceae (plantain family), and Bignoniaceae (trumpet vine family). Extensive research has been undertaken to clarify the interrelationships among these families. Schäferhoff *et al.* used three chloroplast regions, namely trnK/matK, trnL-F, and rps16, for members from all families and successfully resolved the relationships among the nine earliest-branching families with strong support [6]. Refulio-Rodriguez and Olmstead extended the analysis by incorporating seven other gene regions and further refined the phylogeny [7]. Others conducted additional work on Lamiaceae, the largest family, due to its prominence and ecological significance [8-11].

Recent advancements in sequencing technologies, fueled by rapidly decreasing costs, have resulted in the availability of numerous chloroplast sequences in recent years. This progress now makes it possible to compare entire genomes instead of different gene regions. The Pangenome Research Toolkit (PGR-TK) is a cutting-edge tool that utilizes a minimizer-based approach to significantly improve processing speed [12]. This approach enables comprehensive comparisons of entire chloroplast genomes and offers visual representations of similarity blocks, providing valuable insights into the evolutionary relationships among the families studied. The effectiveness of PGR-TK was demonstrated previously [13, 14], and this study applied it to the 87 genera and 14 families within Lamiales for which complete plastid sequences were available.

# Results

Plastid sequences were downloaded from NCBI and split into FASTA files containing sequences from the individual genera. After removing all FASTA files with only one or two genomes, 87 genera were left. Among those, 27 had ten or more sequences. They were Brandisia (10), Jasminum (10), Ligustrum (10), Osmanthus (10), Amphilophium (11), Scrophularia (11), Torenia (11), Veronica (11), Hemiboea (12), Paraboea (12), Fraxinus (13), Olea (13), Adenocalymma (14), Callicarpa (14), Justicia (15), Scutellaria (17), Erythranthe (19), Chionanthus (20), Myoporum (20), Primulina (25), Buddleja (30), Plantago (30), Pedicularis (33), Phlomoides (48), Salvia (59), Isodon (77), Eremophila (185).

The sets with three or more sequences were analyzed using PGR-TK to identify conservation patterns. Additionally, all plastid sequences were separated at the family level, and PGR-TK analysis was performed on 14 Lamiales families with sufficient data. Detailed methods are provided in [14].

## Observations at the Genus Level

PGR-TK analyzes and compares all input sequences, identifying conserved segments across them. It then generates a visual representation of the sequences, with each conserved segment displayed as a rectangle of a distinct color. Additionally, PGR-TK constructs a cladogram to highlight the evolutionary relationships between the genomes. In this analysis, the PGR-TK profiles were visually inspected for length differences and variations in conservation patterns.

In a typical genus-level PGR-TK comparison of plastids, four distinct clusters of segments are identified, representing long single-copy (LSC) regions, short single-copy (SSC) regions, and two inverted repeat (IR) regions. Additionally, the sequences within the same genus typically share similar lengths. PGR-TK can identify IRs and display both segments in the same color. Moreover, the IR regions are more conserved than the SSC and LSC regions, resulting in PGR-TK plots appearing less fragmented in those areas. Thus, those regions are distinctly recognizable in the plots. Occasionally, some sequences are partially inverted at the inverted repeats. This is a well-recognized pattern among plastids, which maintain two isoforms. We disregard such local inversions in our analysis.

Most genera of Lamiales match the expected pattern described above with the following exceptions (Table 1).

(i) *Incarvillea*, *Orobanche*, and *Utricularia* appear the most divergent from the standard pattern. For *Incarvillea*, the sequences were so divergent that the inverted repeats were not identifiable from the plots (Fig. 1).

(iii) *Acanthus*, *Lindernia*, *Jasminum*, *Elsholtzia*, *Echinacanthus*, *Callitriche*, *Barleria* show extensive fragmentation, although the overall quadripartite structure is preserved. The degree of fragmentation in a sequence corresponded strongly to the level of sequence conservation: higher fragmentation indicates lower conservation, while less fragmentation suggests greater similarity between the sequences.

(iii) For *Amphilophium*, *Eremophila*, *Ruella,* and *Plantago*, plastid sequences have two sets of lengths.

(iv) For *Genlisea*, *Barleria,* and *Pedicularis,* one sequence displays a pattern distinct from the rest. Further analysis of the sequence using the method described in [14] suggests that this plant is misclassified.

## Observations at the Family Level

### Lamiaceae

Lamiaceae [8-11], the largest family within Lamiales, includes a wide range of herbs such as mint, sage, and oregano. Over 200 plastid sequences from this family were analyzed using the PGR-TK pipeline. Despite its size and diversity, the family showed remarkably few discrepancies in its PGR-TK profiles. The primary anomaly was the fragmentation of the *Nepeta* genus, which appeared split across multiple small clusters. This pattern may suggest the presence of distinct *Nepeta* clades.

### Oleaceae

Oleaceae is a large family that spans multiple genera [15]. Overall, the sequences show a high degree of similarity, with only minor variations in length (Fig. 2). Visually, they appear highly uniform and generally cluster according to genus, though a few exceptions exist. For example, one *Olea* sequence diverges from the other plastid sequences, and two *Jasminum* sequences cluster separately from the remaining seven. These anomalies may be attributed to structural rearrangements such as inversions in the SSC or other regions.

Despite these outliers, the dataset demonstrates notable consistency. While a few sequences are shorter, most, particularly those in the topmost cluster of approximately 45–50 sequences, are similar in length. This group may represent a distinct sub-family and warrants further investigation. In contrast, the lower portion of the dataset displays a different clustering pattern, suggesting the possible presence of additional sub-families.

### Gesneriaceae

Members of the Gesneriaceae family [16] exhibit a high degree of sequence similarity. The sequences are nearly identical in length, and the inverted repeat regions align closely, suggesting strong conservation within the group. This family includes a diverse range of genera, such as *Anna*, *Primulina*, and *Hemiboea*, with representation from at least 15–16 different genera. Sequences from individuals within the same genus consistently cluster together, indicating clear phylogenetic coherence.

## Acanthaceae

Acanthaceae comprises several genera, including *Justicia*, *Barleria*, *Strobilanthes*, and *Ruellia*. This dataset contained a relatively small number of plastids (40 in total). While the sequences exhibit some similarities in length and inverted repeat regions, they are not identical. Despite these variations, sequences from the same genus generally cluster together, revealing a clear phylogenetic structure. *Barleria siamensis* clusters with *Strobilanthes,* confirming previous determination of incorrect annotation at the genus level analysis.

## Bignoniaceae

Bignoniaceae [17] also contains a large number of sequences that are generally well-conserved. However, a few sequences from *Incarvillea* and *Campsis* show some differences and appear as outgroups within the entire family.. The rest of the genera, including *Emphylophium*, *Adenocalymma*, and *Anemopagma*, share similar sequences, indicating a conserved group. Overall, the family-level plot is highly fragmented with no clear inverted repeats.

## Lentibulariaceae

This family (Fig. 3) includes the genera *Genlisea*, *Utricularia*, and *Pinguicula*. While genus-level plots for *Genlisea* and *Utricularia* appeared complex, the family-level analysis helps clarify some of these inconsistencies. *Genlisea violacea*, which appeared as an outlier at the genus level, clusters closely with *Pinguicula* at the family level. Most sequences from individual genera group together as expected, with the exception of *Utricularia reniformis*, which emerges as an outgroup to the entire family.

## Orobanchaceae

This family-level plot (Fig. 4) is the most complex within Lamiales, possibly due to the presence of several parasitic plants. The sequences are very different lengthwise, and multiple plastids have unique segments. The extremely small arrows add more difficulty to analyzing this family. One example is *Orobanche*, where sequences near the top are highly similar, mainly from *Pedicularis* but also from *Remania* and several smaller genera. In contrast, the bottom part shows shorter sequences, predominantly from *Orobanche* but also from *Epifagus*, *Agonetia*, and *Canofolis*. There’s a clear distinction between the two groups, and the visually distinct sequences at the bottom deserve further scrutiny.

## Plantaginaceae

The sequences within the Plantaginaceae family [18] generally cluster according to their respective genera of Veronica, Plantago, Caltrix, and others. *Veronica* and *Veronicastrum* sequences form a tight, cohesive group, whereas genera such as *Digitalis*, *Penstemon*, and *Resilia* branch into clearly separate clusters. Apart from approximately half of the notably longer *Plantago* sequences, sequence lengths are largely consistent within this family.

## Scrophulariaceae

This family includes the genera *Buddleja*, *Diocirea*, *Eremophila*, *Myoporum*, *Scrophularia*, and *Verbascum*. The sequences across these genera exhibit a high degree of similarity. All genera except *Scrophularia* form cohesive individual clusters in the phylogeny. Interestingly, *Scrophularia* diverges into two distinct groups, suggesting possible substructuring within the genus.

## Linderniaceae

This family consists of only a few genera: *Torenia*, *Lindeira*, *Bonilla*, and *Vandelia*. The sequences in this group show a high degree of similarity, especially in the inverted repeats, and the sequences are generally well conserved.

## Phrymaceae

Phrymaceae is another interesting family, mainly consisting of *Erythranthe* and *Diplacus*. The sequences appear to be mostly conserved, with simple and clear grouping of the genera. Only *Erythranthe parviflora* grouped with *Diplacus,* suggesting possible misclassification of the plant. This species is also the most outlying member in the genus-level plot of *Erythranthe*.

## Verbenaceae

This family includes a small number of sequences from several genera. Though the sequences are reasonably conserved, they are not as uniform as those in the previous families, like Bignoniaceae, and exhibit more fragmentation.

## Mazaceae

The PGR-TK plot includes *Mazus* and two sequences from *Lancia*. These sequences show excellent conservation, both in the inverted repeat and other regions, indicating strong homogeneity.

## Paulowniaceae

The PGR-TK plot of this family consists of the *Paulownia* genus and one additional sequence from *Wightia*. Therefore, the family-level plot does not provide further insight. The *Wightia* sequence appears similar to the *Paulownia* sequences.

# Discussion

In this study, we analyze genus- and family-level plastome structural plots for members of the Lamiales. Broadly speaking, the data reveal consistent patterns across taxa. In a previous paper, we characterized genus-level plastome structures across all angiosperms, and the general patterns observed in Lamiales align well with those earlier findings. Most plots clearly delineate the four major regions of the plastome: the large single-copy (LSC), small single-copy (SSC), and the two inverted repeats (IRs). However, moving to family-level analyses allows for the resolution of certain ambiguities that were less clear at the genus level.

Among the family-level plots, three in particular stand out for their complexity: the Orobanchaceae, Bignoniaceae and Lentibulariaceae. The Orobanchaceae present a particularly intriguing case. This family includes several parasitic species with significantly reduced plastomes. Their small size and structural variation make alignment and visualization more difficult. The family-level plot for Orobanchaceae exhibits substantial complexity, with both length variability and sequence-level divergence. This is consistent with known trends in parasitic lineages, where plastome degradation is common.

Bignoniaceae is another family with a fragmented plastome structure in the family-level plot. Interestingly, although the plastome sequences have roughly similar total lengths, their structural integrity is highly disrupted. This fragmentation appears to be driven by the inclusion of genera such as *Incarvillea* and *Campsis*. As discussed in our previous genus-level analysis, *Incarvillea* in particular displays a highly fragmented PGR-TK plot lacking a recognizable inverted repeat. When such divergent taxa are incorporated into the broader family-level analysis, they contribute significant noise and fragmentation to the overall structural plot.

In case of Lentibulariaceae, which includes genera such as *Genlisea*, *Utricularia*, and *Pinguicula*, the family-level plot resolves ambiguities observed at the genus level. In our genus-level analysis, plastome structures for *Genlisea* and *Utricularia* appeared unusually complex. However, the family-level plot now reveals clearer patterns. For instance, one *Genlisea* plastome that initially appeared highly divergent is now shown to cluster closely with *Pinguicula*, suggesting that the earlier divergence was likely due to a misannotation. In summary, while genus-level analyses provide a good foundation, family-level plastome structure comparisons can uncover both deeper evolutionary patterns and potential issues such as misannotation or structural anomalies. These insights are critical for improving plastome assemblies and for understanding evolutionary dynamics across Lamiales.

# Figures

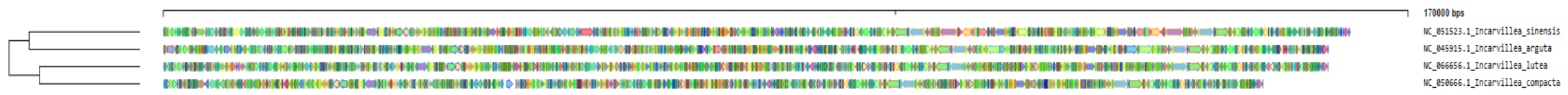

**Figure 1: PGR-TK plot of Incarvillea.** The sequences are so divergent that no clear inverted repeat pattern is visible.

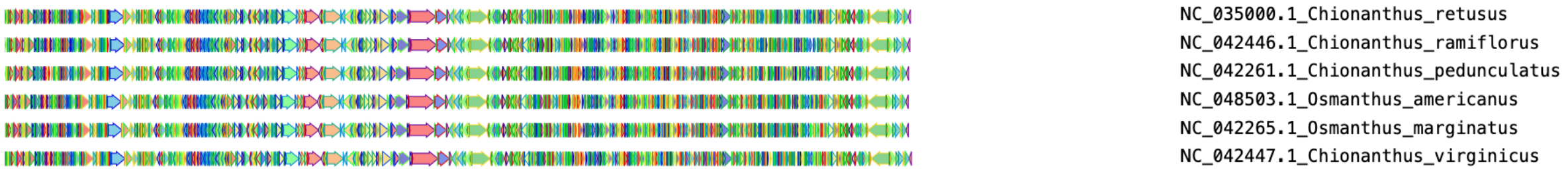

**Figure 2: Partial PGR-TK plot of the Oleaceae family.** The plastids are all highly similar, with only minor length changes.

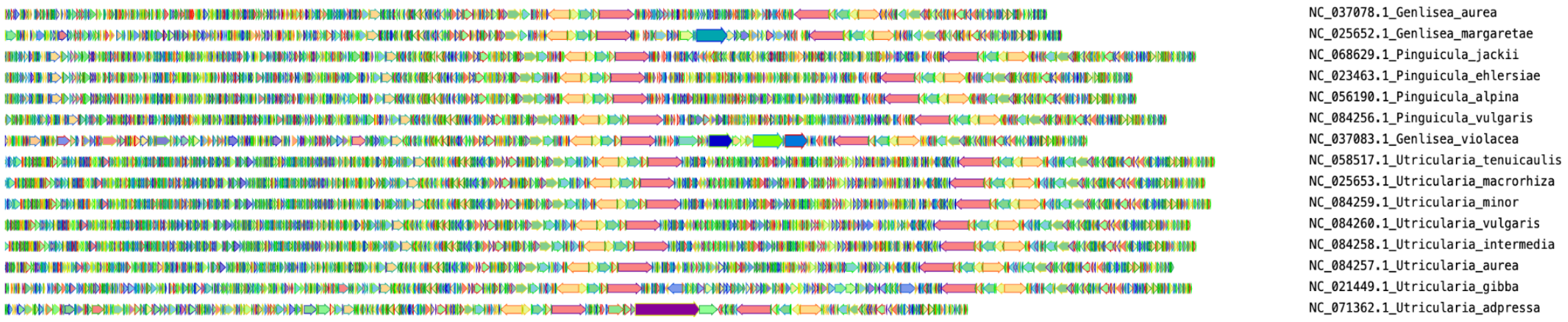

**Figure 3: Partial PGR-TK plot of the Lentibulariaceae family.** The plastids have varying lengths and repeats. Also, *Genlisea violacea* clusters with *Pinguicula*.

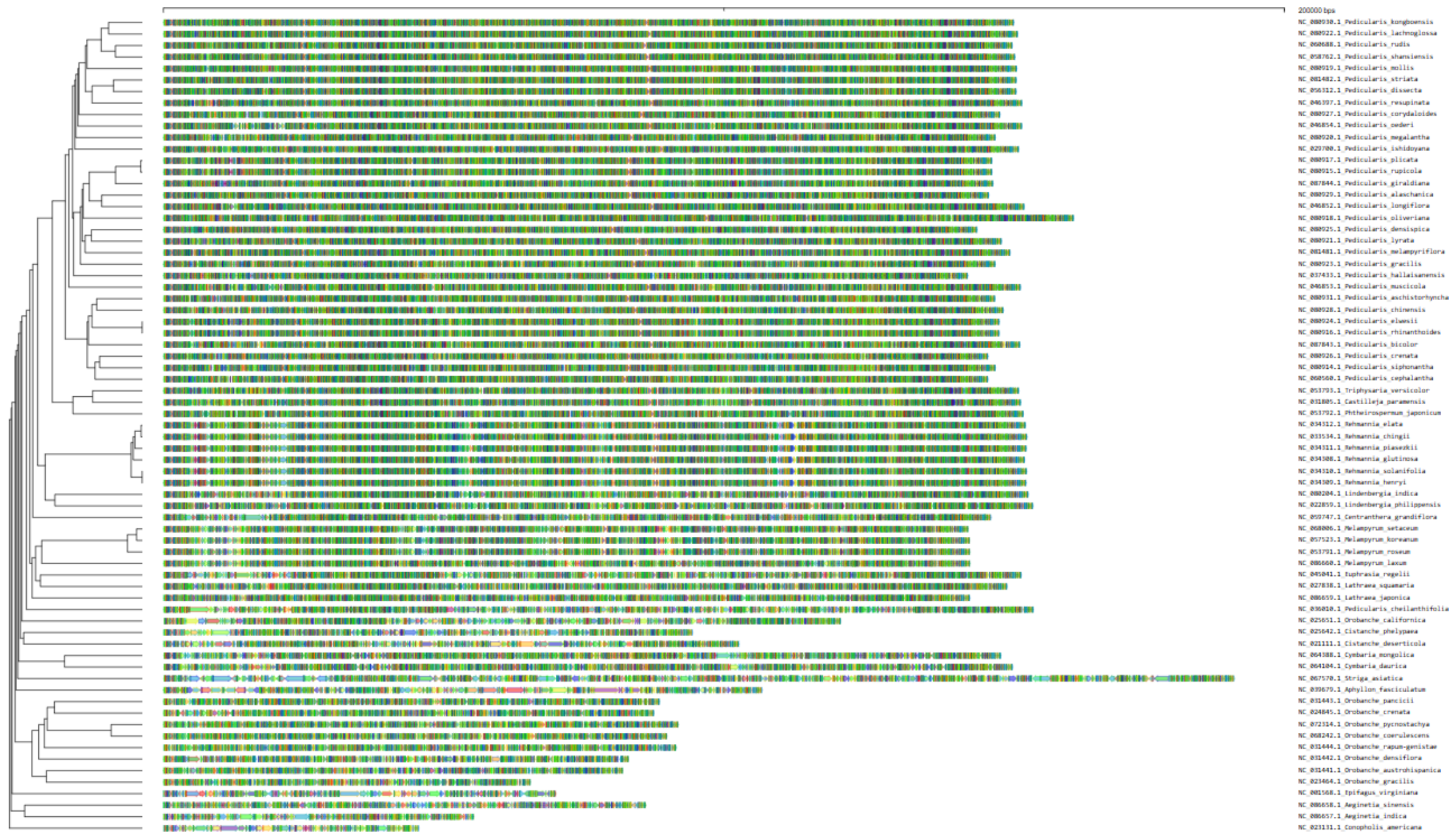

**Figure 4: PGR-TK plot of the Orobanchaceae family.** Multiple distinct sequences near the bottom do not match up with each other, especially in *Orobanche*.

# Tables

| Genus | Family | Pattern Class (based on Ref 14) | Comment |
|---|---|---|---|
| *Acanthus* | Acanthaceae | 2 | |
| *Adenocalymma* | Bignoniaceae | 1 | |
| *Ajuga* | Lamiaceae | 1 | |
| *Amphilophium* | Bignoniaceae | 1 | Two different lengths observed |
| *Anemopaegma* | Bignoniaceae | 1 | |
| *Anna* | Gesneriaceae | 1 | |
| *Avicennia* | Acanthaceae | 1 | |
| *Barleria* | Acanthaceae | 2, 3 | One diverging sequence is close to *Strobilanthes crispa* |
| *Bonnaya* | Linderniaceae | 1 | |
| *Brandisia* | Orobanchaceae | 1 | |
| *Buddleja* | Scrophulariaceae | 1 | |
| *Callicarpa* | Lamiaceae | 1 | |
| *Callitriche* | Plantaginaceae | 2 | |
| *Caryopteris* | Lamiaceae | 1 | |
| *Catalpa* | Bignoniaceae | 1 | |
| *Chionanthus* | Oleaceae | 1 | |
| *Clerodendrum* | Lamiaceae | 1 | |
| *Clinopodium* | Lamiaceae | 1 | |
| *Colquhounia* | Lamiaceae | 1 | |
| *Didymocarpus* | Gesneriaceae | 1 | |
| *Diocirea* | Scrophulariaceae | 1 | |
| *Diplacus* | Phrymaceae | 1 | |
| *Dracocephalum* | Lamiaceae | 1 | |
| *Echinacanthus* | Acanthaceae | 2 | |
| *Elsholtzia* | Lamiaceae | 2 | |
| *Eremophila* | Scrophulariaceae | 1 | Two different lengths observed |
| *Erythranthe* | Phrymaceae | 1 | |
| *Forestiera* | Oleaceae | 1 | |
| *Forsythia* | Oleaceae | 1 | |

| Genus | Family | Pattern Class (based on Ref 14) | Comment |
|---|---|---|---|
| *Acanthus* | Acanthaceae | 2 | |
| *Adenocalymma* | Bignoniaceae | 1 | |
| *Fraxinus* | Oleaceae | 1 | |
| *Genlisea* | Lentibulariaceae | 2, 3 | |
| *Hemiboea* | Gesneriaceae | 1 | |
| *Incarvillea* | Bignoniaceae | 2 | |
| *Isodon* | Lamiaceae | 1 | |
| *Jasminum* | Oleaceae | 2 | One sequence is shorter than the rest |
| *Justicia* | Acanthaceae | 1 | |
| *Leonurus* | Lamiaceae | 1 | |
| *Ligustrum* | Oleaceae | 1 | |
| *Linaria* | Plantaginaceae | 1 | |
| *Lindernia* | Linderniaceae | 2 | |
| *Mazus* | Mazaceae | 1 | |
| *Melampyrum* | Orobanchaceae | 1 | |
| *Melissa* | Lamiaceae | 1 | |
| *Mentha* | Lamiaceae | 1 | |
| *Myoporum* | Scrophulariaceae | 1 | |
| *Nepeta* | Lamiaceae | 1 | |
| *Nestegis* | Oleaceae | 1 | |
| *Noronhia* | Oleaceae | 1 | |
| *Notelaea* | Oleaceae | 1 | |
| *Ocimum* | Lamiaceae | 1 | |
| *Olea* | Oleaceae | 1 | |
| *Oreocharis* | Gesneriaceae | 1 | |
| *Orobanche* | Orobanchaceae | 3, 4 | |
| *Osmanthus* | Oleaceae | 1 | |
| *Paraboea* | Gesneriaceae | 1 | |
| *Paraphlomis* | Lamiaceae | 1 | |
| | | | |
| *Paulownia* | Paulowniaceae | 1 | |
| *Pedicularis* | Orobanchaceae | 2, 3 | One diverging sequence closely matches *Dracocephalum heterophyllum* |
| *Penstemon* | Plantaginaceae | 1 | |
| *Petrocodon* | Gesneriaceae | 1 | |

| Genus | Family | Pattern Class (based on Ref 14) | Comment |
|---|---|---|---|
| *Acanthus* | Acanthaceae | 2 | |
| *Adenocalymma* | Bignoniaceae | 1 | |
| *Phillyrea* | Oleaceae | 1 | |
| *Phlomoides* | Lamiaceae | 1 | |
| *Pinguicula* | Lentibulariaceae | 1 | |
| *Plantago* | Plantaginaceae | 1 | Two different lengths observed |
| *Pogostemon* | Lamiaceae | 1 | |
| *Primulina* | Gesneriaceae | 1 | |
| *Prostanthera* | Lamiaceae | 1 | |
| *Rehmannia* | Orobanchaceae | 1 | |
| *Ruellia* | Acanthaceae | 1 | Two sequences are shorter than the other two |
| *Salvia* | Lamiaceae | 1 | |
| *Schnabelia* | Lamiaceae | 1 | |
| *Schrebera* | Oleaceae | 1 | |
| *Scrophularia* | Scrophulariaceae | 1 | |
| *Scutellaria* | Lamiaceae | 1 | |
| *Stachys* | Lamiaceae | 1 | |
| *Stenogyne* | Lamiaceae | 1 | |
| *Strobilanthes* | Acanthaceae | 1 | |
| *Syringa* | Oleaceae | 1 | |
| *Tanaecium* | Bignoniaceae | 1 | |
| *Thymus* | Lamiaceae | 1 | |
| *Torenia* | Linderniaceae | 1 | |
| *Utricularia* | Lentibulariaceae | 5 | |
| *Verbascum* | Scrophulariaceae | 1 | |
| *Verbena* | Verbenaceae | 1 | |
| *Veronica* | Plantaginaceae | 1 | |
| *Veronicastrum* | Plantaginaceae | 1 | |
| *Vitex* | Lamiaceae | 1 | |

**Table 1. PGR-TK Patterns of Lamiales Genera.** This table summarizes the patterns observed among various PGR-TK plots from 87 genera of Lamiales. The patterns are described in Ref [14]. Additional comments are provided for some genera.

# References

1. Lamiales [Internet]. [cited 2025 Apr 29]. Available from: https://www.mobot.org/MOBOT/research/APweb/orders/lamialesweb.htm#Lamiales

2. An update of the Angiosperm Phylogeny Group classification for the orders and families of flowering plants: APG IV. Botanical Journal of the Linnean Society. 2016 Mar 24;181(1):1–20.

3. Fonseca LHM. Combining molecular and geographical data to infer the phylogeny of Lamiales and its dispersal patterns in and out of the tropics. Molecular Phylogenetics and Evolution. 2021 Nov;164:107287.

4. Li HT, Yi TS, Gao LM, Ma PF, Zhang T, Yang JB, et al. Origin of angiosperms and the puzzle of the Jurassic gap. Nature Plants. 2019 May 6;5(5):461–70.

5. Alawfi MS, Alzahrani DA. Insights into the phylogenetic relationship of the lamiids orders based on whole chloroplast genome sequencing. Journal of King Saud University - Science. 2023 Jan;35(1):102398.

6. Schäferhoff B, Fleischmann A, Fischer E, Albach DC, Borsch T, Heubl G, et al. Towards resolving Lamiales relationships: Insights from rapidly evolving chloroplast sequences. BMC Evolutionary Biology. 2010;10(1):352.

7. Refulio‐Rodriguez NF, Olmstead RG. Phylogeny of lamiidae. American Journal of Botany. 2014 Feb;101(2):287–99.

8. Li B, Cantino PD, Olmstead RG, Bramley GLC, Xiang CL, Ma ZH, et al. A large-scale chloroplast phylogeny of the Lamiaceae sheds new light on its subfamilial classification. Scientific Reports. 2016 Oct 17;6(1).

9. Zhao F, Chen YP, Salmaki Y, Drew BT, Wilson TC, Scheen AC, et al. An updated tribal classification of Lamiaceae based on plastome phylogenomics. BMC Biology. 2021 Jan 8;19(1).

10. Zhao F, Li B, Drew BT, Chen YP, Wang Q, Yu WB, et al. Leveraging plastomes for comparative analysis and phylogenomic inference within Scutellarioideae (Lamiaceae). PLOS ONE. 2020 May 7;15(5):e0232602.

11. Nazar N, Howard C, Slater A, Sgamma T. Challenges in Medicinal and Aromatic Plants DNA Barcoding—Lessons from the Lamiaceae. Plants. 2022 Jan 5;11(1):137.

12. Chin CS, Behera S, Khalak A, Sedlazeck FJ, Wagner J, Zook JM. Multiscale Analysis of Pangenome Enables Improved Representation of Genomic Diversity For Repetitive And Clinically Relevant Genes [Internet]. Cold Spring Harbor Laboratory; 2022 Aug [cited 2025 Apr 29]. Available from: https://doi.org/10.1101/2022.08.05.502980

13. Jayanti R, Kim A, Pham S, Raghavan A, Sharma A, Samanta MP. arXiv.org. 2023.

Comparative analysis of plastid genomes using pangenome research toolkit (PGR-TK). Available from: https://arxiv.org/abs/2310.19110
14. Samanta MP. arXiv.org. 2025. Pan-genome Analysis of Angiosperm Plastomes using PGR-TK. Available from: https://arxiv.org/abs/2504.20034
15. Hong-Wa C, Besnard G. Intricate patterns of phylogenetic relationships in the olive family as inferred from multi-locus plastid and nuclear DNA sequence analyses: A close-up on Chionanthus and Noronhia (Oleaceae). Molecular Phylogenetics and Evolution. 2013 May;67(2):367–78.
16. Roalson EH, Roberts WR. Distinct Processes Drive Diversification in Different Clades of Gesneriaceae. Systematic Biology. 2016 Feb 14;65(4):662–84.
17. Nazareno AG, Carlsen M, Lohmann LG. Complete Chloroplast Genome of Tanaecium tetragonolobum: The First Bignoniaceae Plastome. PLOS ONE. 2015 Jun 23;10(6):e0129930.
18. Stettler JM, Stevens MR, Meservey LM, Crump WW, Grow JD, Porter SJ, et al. Improving phylogenetic resolution of the Lamiales using the complete plastome sequences of six Penstemon species. PLOS ONE. 2021 Dec 15;16(12):e0261143.